\documentclass[letterpaper, 10 pt, conference]{ieeeconf}  

\IEEEoverridecommandlockouts                              

\usepackage{graphics} 
\usepackage{epsfig} 
\usepackage{amsmath} 
\usepackage{amssymb}  
\usepackage{mathrsfs}
\usepackage{stmaryrd} 
\usepackage{multirow} 

\usepackage[pagebackref=true,breaklinks=true,letterpaper=true,colorlinks,bookmarks=false,citecolor=blue,linkcolor=blue]{hyperref}
\usepackage{etoolbox}
\makeatletter
\patchcmd{\@maketitle}{\raggedright}{\centering}{}{}
\patchcmd{\@maketitle}{\raggedright}{\centering}{}{}
\makeatother

\makeatletter
\let\NAT@parse\undefined
\makeatother

\usepackage{etoolbox}
\makeatletter
\patchcmd{\@makecaption}
  {\scshape}
  {}
  {}
  {}
\makeatletter
\patchcmd{\@makecaption}
  {\\}
  {.\ }
  {}
  {}
\makeatother

\usepackage{caption}

\usepackage{ifthen}
\usepackage{color}
\definecolor{gray}{rgb}{0.5,0.5,0.5}
\definecolor{green}{rgb}{0, 0.4, 0}
\definecolor{orange}{rgb}{1, 0.5, 0}
\definecolor{mahogany}{rgb}{0.75, 0.25, 0.0}
\definecolor{purple}{rgb}{0.6, 0, 0.6}
\definecolor{darkgreen}{rgb}{0, 0.4, 0}
\definecolor{frenchblue}{rgb}{0.0, 0.45, 0.73}
\definecolor{red}{rgb}{1,0,0}
\definecolor{yellow}{rgb}{1,1,0}
\definecolor{magenta}{rgb}{1,0,1}
\title{\LARGE \bf
Plug-and-Play: Improve Depth Prediction via Sparse Data Propagation
}

\begin{document}


\author{Tsun-Hsuan Wang$^{1}$, Fu-En Wang$^{1}$, Juan-Ting Lin$^{1}$, Yi-Hsuan Tsai$^{2}$, Wei-Chen Chiu$^{3}$, Min Sun$^{1}$
\thanks{$^{1}$ National Tsin Hua University}
\thanks{$^{2}$ NEC Labs America}
\thanks{$^{3}$ National Chiao Tung University}}

\maketitle
\thispagestyle{empty}
\pagestyle{empty}

\begin{abstract}

We propose a novel plug-and-play (PnP) module for improving depth prediction with taking arbitrary patterns of sparse depths as input. Given any pre-trained depth prediction model, our PnP module updates the intermediate feature map such that the model outputs new depths consistent with the given sparse depths. Our method requires no additional training and can be applied to practical applications such as leveraging both RGB and sparse LiDAR points to robustly estimate dense depth map. Our approach achieves consistent improvements on various state-of-the-art methods on indoor (i.e., NYU-v2) and outdoor (i.e., KITTI) datasets. Various types of LiDARs are also synthesized in our experiments to verify the general applicability of our PnP module in practice.

\end{abstract}
\section{Introduction}

Depth perception is of great importance in a wide range of applications, including  autonomous driving, robotics, augmented reality, and scene reconstruction. While numerous depth sensing technologies are available, e.g., RGB-D camera, stereo camera rig, and LiDAR, they all have their own pros and cons in different scenarios. For instance, RGB-D or stereo cameras could provide depth estimation which is dense but typically fails on highly reflective and transparent surfaces, or distant objects from the camera. Oppositely, expensive depth sensors such as LiDAR are able to function well under challenging weather conditions and robustly estimate depth of distant objects, but only limit to sparse coverage (e.g., Velodyne 64-channel LiDAR only covers around 5\% of the image space after re-projection~\cite{kitti2017}). Therefore, there is a clear need to take advantage of strengths from different sensing approaches to improve depth perception. In this paper, we particularly tackle the problem of improving dense depth prediction with sparse yet accurate data.

On the one hand, building upon monocular or binocular depth prediction methods, several existing methods (detailed in Sec.~\ref{sec:related}) take sparse depth information as additional input~\cite{sparse-to-dense}\cite{late-fusion}\cite{magic} or propagate sparse data towards their neighborhood~\cite{deep-compressed-sensing}\cite{cspn}. However, these approaches are either inflexible due to requirement of additional training, or being constrained to specific input types, patterns, or density. Hence, they are less practical for real-world applications.

On the other hand, network visualization~\cite{deep-visual} and adversarial attack~\cite{fast-grad} investigate how intermediate feature representations of a network influence the prediction and how such representations can be modified to achieve desirable outputs.
Inspired by these techniques, we introduce a plug-and-play (PnP) module for improving dense depth prediction, which requires no additional training and can be adapted to various input types.
Given a pre-trained depth prediction model, our PnP module updates intermediate feature maps of the original dense depth prediction, enforcing the model to produce a new depth output which is consistent with the given sparse data. Thereby, the proposed PnP module is not only easily integrable to all differentiable depth prediction approaches but also training-free, simply performed during inference. 

We investigate the integration of our PnP module to numerous depth prediction approaches and conduct experiments on both the indoor and outdoor benchmark datasets, i.e., NYU-v2~\cite{nyu-v2} and KITTI~\cite{kitti2012}, respectively. For all the methods on both datasets, our PnP module provides consistent improvements for depth prediction. Moreover, we show the robustness of the proposed method under various settings, including different density and input types. Most importantly, we synthesized various types of LiDARs to verify the general applicability of our method to leverage RGB and sparse LiDAR depth to robustly estimate dense depths. Finally, we demonstrate how our module propagates information from sparse data and improves prediction in the local region, and provide a rule of thumb of using our PnP module via a thorough ablative analysis. The source code is available at \url{https://zswang666.github.io/PnP-Depth-Project-Page/}.

\begin{figure}[t!]
\centering
\includegraphics[width=0.44\textwidth]{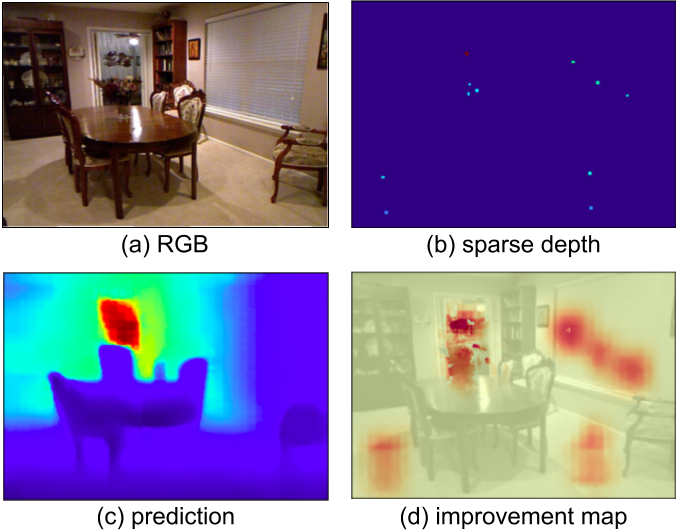}
\vspace{-2mm}
\caption{\small{Given an (a) RGB image and a depth estimation/completion model, our PnP module propagates information from (b) sparse points and generates (c) improved depth prediction without any re-training. We show the (d) improvement denoted in red region.}}
\label{fig:teaser}
\vspace{-6mm}
\end{figure}
\section{Related Work}\label{sec:related}


{\flushleft {\bf RGB-based Depth Estimation.}}
Monocular or binocular depth estimation has been studied widely in robotics and computer vision.
Recent works are mostly built upon deep networks with differentiable architectures.
For instance, \cite{eigen1}\cite{eigen2} first propose a multi-scale network to predict depth map in a coarse-to-fine manner. In order to increase the edge response on the estimated depth map, Liu \textit{et al.} \cite{conv-neural-field} develop a deep convolutional neural field model to incorporate continuous Conditional Random Field (CRF) technique into the network. The performance of depth estimation is further boosted by Laina \textit{et al.} \cite{fcrn} via introducing a powerful residual network and up-projection layers in the deep architecture. More recently, Fu \textit{et al.} \cite{dorn} propose to formulate depth prediction as an ordinal regression problem and obtain impressive results on depth estimation.
Despite that these depth estimation methods use various network architectures,
our PnP module can be flexibly integrated into these approaches without re-training their models and requiring specific constraints on the network choice.


\vspace{-1mm}
{\flushleft {\bf Depth Reconstruction from Sparse Data.}}
  The nature of obtained high-precision but sparse depth maps from active depth sensors (e.g., LiDAR, Kinect) recently brings the research interests in sparse-to-dense depth reconstruction and completion. \cite{hawe-wavelet}\cite{liu-wavelet} perform depth reconstruction upon hypothesizing the sparsity of disparity maps in the Wavelet domain. Ferstl \textit{et al.} \cite{tgv-anisotropic} propose to upsample low-resolution depth maps guided by anisotropic diffusion tensors. For specific sensors, Uhrig \textit{et al.} \cite{kitti2017} design a sparse convolution layer to handle missing data captured from LiDAR scans, and Liao \textit{et al.} \cite{parse-geometry} construct a residual network to parse the 2D planar measurements obtained by a laser range finder. In order to design a more generalizable method not only for particular sensors, Ma \textit{et al.} \cite{sparse-to-dense} and Jaritz \textit{et al.} \cite{late-fusion} respectively adopt early-fusion and late-fusion strategies to combine the RGB image and sparse depth for producing the final depth estimation. \cite{deep-depth-completion} further studies the connection between depth and surface normal maps for depth completion. In addition, Chodosh \textit{et al.} \cite{deep-compressed-sensing} combine compressed sensing with deep learning into a single framework and perform depth completion to be consistent with the given sparse data. This model shares a similar idea to ours of optimizing intermediate representation to yield the depth prediction.

While the above-mentioned works propose stand-alone approaches for depth reconstruction, our goal is to design a module that can be flexibly integrated into any existing methods and improve their performance.
From this modeling perspective, the target task is extended from depth reconstruction further to depth refinement, where the initial result of depth estimation or reconstruction is boosted by the given sparse data. 
To the best of our knowledge, the work closest to our scope is \cite{cspn} that proposes a convolutional spatial propagation network by concatenating it to a depth estimation network thus achieves depth refinement. However, this approach~\cite{cspn} requires an additional training stage while our method simply needs to perform the inference stage. It is also worth noting that, with the given sparse data, our method can be comfortably applied to both depth reconstruction and depth refinement approaches. 

\section{Method}

\begin{figure*}[t!]
\centering
\includegraphics[width=0.7\textwidth]{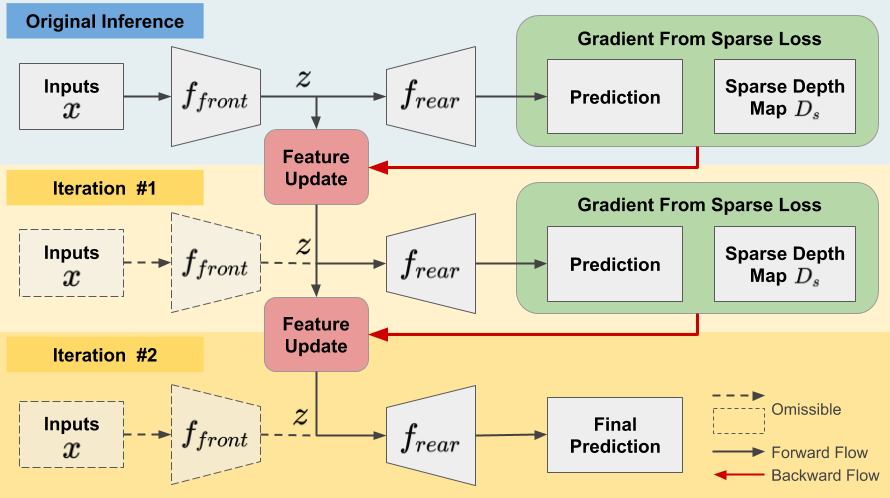}
\vspace{-1mm}
\caption{\small{\textbf{Illustration of the PnP module via backward-forward propagation.} Given a depth estimation model $f$ and sparse ground truth $D_s$, our PnP module iteratively updates the intermediate representation $z$ based on the gradient computed from sparse points and re-inference to obtain better depth prediction. Here, we illustrate an example for two iterations.}}
\label{fig:overview}
\vspace{-5mm}
\end{figure*}

Different from existing depth reconstruction methods that design a full model \cite{sparse-to-dense}\cite{late-fusion}\cite{parse-geometry}\cite{deep-depth-completion} or a refinement component \cite{deep-compressed-sensing}\cite{cspn} that requires additional training, we propose a plug-and-play module that is used during inference, without updating model weights, and can be applied to any differentiable models, including the above-mentioned methods and existing depth estimation approaches. 

Given the input data $x$ (e.g., an RGB image) and a depth estimation model $f$, our goal is to improve the dense depth prediction $f(x)$ with the help of the sparse depth $D_s$ during inference. Here, input $x$ could be RGB images, sparse depths, or both. 
In the following, we sequentially introduce the proposed PnP module, demonstrate its usefulness with mathematical insights, and describe a rule of thumb of utilizing our module.

\vspace{-1mm}
\subsection{Preliminary for Depth Estimation}{
Without loss of generality, the objective of learning-based approaches for depth estimation can be described as follows. 
Given the input data $x$, we aim to minimize the error between the prediction $f(x)$ and the ground truth depth $D$, with respect to the network $f$ parametrized by $\theta$:
%
\begin{equation}
	\theta^* = \underset{\theta}{\arg\min}~\mathcal{L}(f(x; \theta), D),
    \label{eq: obj_theta}
\end{equation}
where $\mathcal{L}(\cdot,\cdot)$ is the loss function that could be mean absolute error ($L_1$), mean squared error ($L_2$), reversed Huber loss \cite{fcrn}, inverse depth loss \cite{late-fusion}, or classification loss \cite{dorn}\cite{parse-geometry}\cite{cao}.

Although the predicted depth $f(x; \theta)$ can be affected by both the model parameters $\theta$ and the input $x$, typically only the parameters $\theta$ are updated in learning depth estimation network. In our method, we aim to design a plug-and-play module for improving the depth estimation without any additional training, such that the $\theta$ of network $f$ needs to remain fixed whereas updating $x$ seems the only feasible way in order to minimize the $\mathcal{L}$ loss. This idea is analogous to the one in adversarial attack~\cite{fast-grad} (which we are motivated from), where the network is unchanged and the perturbation is added to input $x$ for making the new prediction towards an assigned target output. 
In practice, instead of directly updating $x$, we choose to update an intermediate feature representation that is a function of $x$ for efficiency (details described as follows). 
}

\vspace{-1mm}
\subsection{Sparse Data Propagation}{
%
%
%
Since we aim to update the intermediate feature, which corresponds to the response of a certain layer in depth estimation network, the network $f$ can be treated as a cascade of two parts, $f_{front}$ and $f_{rear}$, such that $z=f_{front}(x)$ and $f(x)=f_{rear}(f_{front}(x))$. The essence of our method is to update $z$ in the direction that improves depth prediction $f(x)$ with partial ground truth $D_s$ and it can be formulated as:
\begin{equation}
	\begin{split}
		z^* &= \underset{z}{\arg\min}~\mathcal{L}(f_{rear}(z; \theta_{rear}), D_s), \\
    	f(x) &= f_{rear}(z^*),
    \end{split}
    \label{eq:objective}
\end{equation}
where the value of $z$ during minimization is initialized as $f_{front}(x)$. Similar to training a deep network, we adopt back-propagation \cite{back-prop} to solve the minimization problem over $z$. The iterative optimization process is written as:
\begin{equation}
	\begin{split}
    	z_0 &= f_{front}(x), \\
        z_{k+1} &= z_k - \alpha U(\frac{\partial \mathcal{L}(f_{rear}(z_k), D_s)}{\partial z_k}),
    \end{split}
    \label{eq:iter}
\end{equation}
where $U(\cdot)$ can be any variants of gradient descent methods such as Adam~\cite{adam} and $\alpha$ is the learning rate. For simplicity, we adopt fast gradient sign method~\cite{fast-grad}, where $U(\cdot)=\textit{sign}(\cdot)$ that aligns well with the adversarial attack technique used in~\cite{fast-grad} and is found to be both efficient and empirically effective in experiments.
Fig.~\ref{fig:overview} illustrates how our proposed PnP module works. By alternatingly performing back-propagation and re-inference on the depth estimation network, which we denote this procedure as \textit{backward-forward propagation}, the information of the given sparse depth is propagated to their corresponding local regions. We also note that, the parameters of network $f$ is not updated but only $z$ is.
%
%
In addition, the choice of which intermediate layer for $z$ depends on whether it contains sufficiently large area impacted in the output depth map, which we name this as \textit{influential field} and we will discuss it in Sec.~\ref{ssec:how_to_use}.
}

\vspace{-1mm}
\subsection{Theoretical Discussion}{
\label{ssec:insights}
We have described the proposed PnP module to update the intermediate feature map $z$ based on the sparse points only. However, it still remains a question whether such updates guarantee improvement in dense depth map.
In this section, we aim to demonstrate that our method provides sufficiently correct guidance for dense depth reconstruction. Here, we first consider the gradient with respect to $z$ if the dense ground truth $D$ is given:
\begin{equation}
	g_{ij} = \frac{\partial \mathcal{L}(f_{rear}(z)_{ij},D_{ij})}{\partial z},
    \label{eq:g_ij}
\end{equation}
where $g_{ij}$ provides a gradient for $z$ toward correct prediction guided by the given ground truth at pixel $(i,j)$. Assuming $M$ is a mask such that $D_s=M\cdot D$, we can compute the gradient based on the sparse ground truth $D_s$:
\begin{equation}
	\begin{split}
	\hat{g}_{ij} &= \frac{\partial M_{ij}\mathcal{L}(f_{rear}(z)_{ij},D_{ij})}{\partial z} \\
    &= M_{ij}g_{ij} + \frac{\partial M_{ij}}{\partial z}\mathcal{L}(f_{rear}(z)_{ij},D_{ij}),
    \end{split}
    \label{eq:ghat_ij}
\end{equation}
where $M_{ij}$ is $1$ if the depth is given and $0$, otherwise. Comparing \eqref{eq:g_ij} and \eqref{eq:ghat_ij}, $\hat{g}$ is a masked $g$ plus a residual term.
%
For the residual term, when $D_s$ is included in the input $x$, $z$ is a function of $D_s$ as $z=f_{front}(x)$, resulting in $z$ also being a function of $M$, and thus this term is not always zero.
%
%

There are two cases for the residual term being a small value, which indicates that the gradient computed using sparse points still provides an approximately correct direction for the dense map: 1) $M_{ij}=1$ when sparse ground truth is provided on pixel $(i,j)$, leading to small $\mathcal{L}(f_{rear}(z)_{ij},D_{ij})$; and 2) $M_{ij}=0$ when sparse ground truth is not provided on pixel $(i,j)$, in which $z$ should not be affected by $M_{ij}$, resulting in $\frac{\partial M_{ij}}{\partial z}\approx 0$.
}

\vspace{-1mm}
\subsection{Property and Usage}{
\label{ssec:how_to_use}
In this section, we describe the property of the proposed PnP module and how it is applicable to existing models.

\vspace{-1mm}
{\flushleft {\bf Influential Field.}}
We define the local area propagated from the sparse points as the influential field, in which this field is confined by the backward-forward propagation process. The idea of influential field is similar to the one of receptive field, while the former indicates a region in the network output and the latter points to the input. 
For example, back-propagating through two consecutive $3\times 3$ convolution layers with stride $1$ yields a $5\times 5$ influential field in our method.
As such, in order to enlarge the propagated area, our module suggests to choose an intermediate feature map $z$ that covers sufficiently large influential field. Please note that \cite{cspn} formulates depth refinement as an anisotropic diffusion process and the information of sparse points is iteratively diffused to their neighboring regions, in which its concept is similar to our influential field but needs additional training.


\vspace{-1mm}
{\flushleft {\bf Number of Iteration.}}
During the process of backward-forward propagation, we can control how many iterations are performed. With more iterations, our module is able to gradually update the dense depth map from the sparse points but result in slower inference time. We will show in the experimental section that the improvement gradually saturates as iteration grows, and hence large iteration number is not preferable.

}

\section{Experimental Results}
\label{sec: exp}
In this section, we first introduce the baseline methods used to integrate with the proposed PnP module and the datasets for evaluation.
Second, we present experimental results of our method with various types of input. Further analysis on the property of our PnP module is also presented.

\vspace{-1mm}
\subsection{Baselines and Implementation Details}
With the given sparse points of depth, our PnP module can further improve the depth estimation during inference. Here, we select numerous representative approaches for RGB-based depth estimation and depth reconstruction.

\begin{table*}[t]
\centering
\caption{\small{Evaluation on NYU-Depth-v2. Uniform sampling is adopted to obtain sparse samples following \cite{sparse-to-dense}.}}
\begin{tabular}{l l l l l l l l l l l}
\hline
Problem & Method & \#Samples & \%Samples & RMSE (m)~$\shortdownarrow$ & MAE (m)~$\shortdownarrow$ & MRE~$\shortdownarrow$ & $\delta_1\shortuparrow$ & $\delta_2\shortuparrow$ & $\delta_3\shortuparrow$ \\ \hline \hline
\multirow{4}{*}{rgb} & Eigen \textit{et al.}~\cite{eigen1} & - & - & $0.8933$ & $0.7076$ & $0.2862$ & $52.1$ & $82.3$ & $93.8$ \\
					 & Our + Eigen         & $100$ & $0.145$ & $0.5021 (+43.8\%)$ & $0.3765 (+46.8\%)$ & $0.1497$ & $80.5$ & $94.8$ & $98.8$ \\
                     & FCRN~\cite{fcrn}    & - & - & $0.5111$ & $0.3665$ & $0.1369$ & $81.5$ & $95.4$ & $98.8$ \\
                     & Our + FCRN          & $100$ & $0.145$ & $0.3720 (+27.2\%)$ & $0.2438 (+33.5\%)$ & $0.0890$ & $91.5$ & $98.3$ & $99.6$ \\
\hline
\multirow{2}{*}{sd} & Uhrig \textit{et al.}~\cite{kitti2017} & $3466$ & $5$ & $0.1154$ & $0.0500$ & $0.0177$ & $99.3$ & $99.9$ & $99.9$ \\
					& Our + Uhrig            & $3466$ & $5$ & $0.1050 (+8.9\%)$ & $0.0441 (+11.8\%)$ & $0.0162$ & $99.4$ & $99.9$ & $99.9$ \\
\hline
\multirow{4}{*}{rgb + sd} & Ma \textit{et al.}~\cite{sparse-to-dense} & $500$ & $0.725$ & $0.1982$ & $0.1125$ & $0.0398$ & $97.9$ & $99.6$ & $99.9$ \\
					      & Our + Ma              & $500$ & $0.725$ & $0.1782 (+10.1\%)$ & 0.0087 (+22.7\%) & $0.0315$ & $98.2$ & $99.7$ & $99.9$ \\
                          & Jaritz \textit{et al.}~\cite{late-fusion} & $500$ & $0.725$ & $0.1712$ & $0.0996$ & $0.0343$ & $98.7$ & $99.7$ & $99.9$ \\
                          & Our + Jaritz           & $500$ & $0.725$ & $0.1598 (+6.7\%)$ & $0.0817 (+17.9\%)$ & $0.0320$ & $98.7$ & $99.7$ & $99.9$ \\
\hline
\end{tabular}
\vspace{-2mm}
\small
\label{tab:baselines-nyu}
\end{table*}

\begin{table*}[t]
\centering
\caption{\small{Evaluation on KITTI. $\dag$ denotes the KITTI \textit{Depth Completion} dataset and the others are evaluated on the KITTI \textit{Odometry} dataset. Uniform sampling is adopted to obtain sparse samples on KITTI \textit{Odometry} following \cite{sparse-to-dense}.}}
\begin{tabular}{l l l l l l l l l l}
\hline
Problem & Method & \#Samples & \%Samples & RMSE (m)~$\shortdownarrow$ & MAE (m)~$\shortdownarrow$ & MRE~$\shortdownarrow$ & $\delta_1\shortuparrow$ & $\delta_2\shortuparrow$ & $\delta_3\shortuparrow$ \\ \hline \hline
\multirow{4}{*}{rgb} & Eigen \textit{et al.}~\cite{eigen1} & - & - & $6.6128$ & $3.9234$ & $0.2249$ & $58.9$ & $86.4$ & $94.7$ \\
                     & Our + Eigen & $500$ & $0.5$ & $5.1395 (+22.3\%)$ & $2.8633 (+27.0\%)$ & $0.1790$ & $75.1$ & $91.7$ & $97.0$ \\
                     & Zhou \textit{et al.}~\cite{sfm-learner} & - & - & $7.2722$ & $4.1411$ & $0.2895$ & $63.6$ & $85.7$ & $93.3$ \\
                     & Our + Zhou & $500$ & $0.1$ & $6.0726 (+16.5\%)$ & $3.4042 (+17.8\%)$ & $0.2269$ & $71.0$ & $89.6$ & $95.2$ \\ 
\hline
\multirow{4}{*}{sd} & Uhrig \textit{et al.}~\cite{kitti2017} & \multicolumn{2}{l}{LiDAR data ($\sim 5\%)^{\dag}$} & $1.6616$ & $0.5188$ & $0.0311$ & $39.0$ & $66.9$ & $81.6$ \\
					& Our + Uhrig            & \multicolumn{2}{l}{LiDAR data ($\sim 5\%)^{\dag}$} & $1.6244 (+2.2\%)$ & $0.4533 (+12.6\%)$ & $0.0275$ & $50.7$ & $75.7$ & $86.1$ \\
                    & Chodosh \textit{et al.}~\cite{deep-compressed-sensing} & \multicolumn{2}{l}{LiDAR data ($\sim 5\%)^{\dag}$} & $1.3859$ & $0.4813$ & $0.0276$ & $34.7$ & $64.4$ & $80.0$ \\
                    & Our + Chodosh          & \multicolumn{2}{l}{LiDAR data ($\sim 5\%)^{\dag}$} & $1.3602 (+1.9\%)$ & $0.4192 (+12.9\%)$ & $0.0233$ & $52.7$ & $75.8$ & $85.1$ \\
\hline
\multirow{2}{*}{rgb + sd} & Ma \textit{et al.}~\cite{sparse-to-dense} & $500$ & $0.1$ & $3.0272$ & $1.1879$ & $0.0703$ & $94.8$ & $97.9$ & $99.0$ \\
                          & Our + Ma & $500$ & $0.1$ & $2.9754 (+1.7\%)$ & $1.0241 (+13.79\%)$ & $0.0579$ & $94.9$ & $98.0$ & $99.0$ \\
\hline
\end{tabular}
\small
\label{tab:baselines-kitti}
\vspace{-5mm}
\end{table*}

\vspace{-1mm}
{\flushleft {\bf RGB-based Depth Estimation Methods.}}
Two commonly used monocular depth estimation methods~\cite{eigen1}\cite{fcrn} and one unsupervised approach~\cite{sfm-learner} are considered:
\begin{itemize}
\item \textbf{Eigen \textit{et al.}}~\cite{eigen1} utilize a coarse network of depth estimation followed by a refinement model that takes the coarse output and RGB image as inputs to produce the final depth estimation. Since the refinement network is only composed of two convolutional layers, we further back-propagate through the coarse network;
\item \textbf{FCRN}~\cite{fcrn} is built upon ResNet-50 followed by up-sample blocks to decode features for dense estimation. We back-propagate gradients to the layer before the first up-sample block;
\item \textbf{Zhou \textit{et al.}}~\cite{sfm-learner} adopt an U-Net based model and we select the bottleneck feature map as the intermediate feature map $z$, while all the skip-connected features are not updated.
\end{itemize}


\vspace{-1mm}
{\flushleft {\bf Depth Reconstruction Methods.}}
For the approaches that already use the sparse depth as the input, our PnP module can still improve the estimation.
Here, we consider two depth reconstruction methods~\cite{kitti2017}\cite{deep-compressed-sensing} that take sparse depth only as the input and the other two algorithms~\cite{sparse-to-dense}\cite{late-fusion} that input both the sparse depth and RGB image.
\begin{itemize}
\item \textbf{Uhrig \textit{et al.}}~\cite{kitti2017} design a sparse convolutional operation to reconstruct dense depth from sparse LiDAR data. We choose the output of the first sparse convolutional layer as the intermediate feature map $z$ for sufficiently large influential field;
\item \textbf{Chodosh \textit{et al.}}~\cite{deep-compressed-sensing} develop a feed-forward technique to obtain an optimal $x$ such that the predicted depth map $f(x)$ is consistent with the given sparse depth $D_s$. There are only three layers in their network $f$ and we select the optimal $x$ as the intermediate feature map $z$;
%
\item \textbf{Ma \textit{et al.}}~\cite{sparse-to-dense} adopt a ResNet-based encoder-decoder network to predict dense depth map from the concatenation of an RGB image and a sparse depth map. We select the bottleneck feature map to be the intermediate representation $z$ for back-propagation;
\item \textbf{Jaritz \textit{et al.}}~\cite{late-fusion} introduce late fusion for sparse depth data and predict dense depth map through a skip-connected encoder-decoder network. We empirically found that the output is not sensitive to the bottleneck feature map (32x) and thus we select a higher resolution feature map (8x) as the intermediate representation $z$.
\end{itemize}

\vspace{-1mm}
{\flushleft {\bf Implementation Details.}}
In our PnP module, the learning rate $\alpha$, the number of iterations, and the loss function are set as 0.01, 5, and $L_1$ loss respectively for all experiments if not particularly mentioned.
To evaluate the improvement introduced by our method with fair comparisons, all the baselines follow the original settings (e.g., number of sparse samples and sampling strategy) as mentioned in their original papers. For instance, as some baselines adopt uniform sampling to obtain sparse depth, we then use the same random seed as theirs to create sparse data for our PnP module.

\vspace{-1mm}
\subsection{Datasets and Metrics}
\vspace{-1mm}
{\flushleft {\bf NYU-Depth-v2 Dataset.}}
The NYU-Depth-v2 dataset~\cite{nyu-v2} consists of RGB and depth images captured from a Microsoft Kinect in 464 different indoor scenes. The official split contains 249 training and 215 testing scenes. For benchmarking, we compute error measures on the commonly used test subset of 654 images, following~\cite{sparse-to-dense}\cite{fcrn}. The depth images in the dataset are reprojected to the RGB image coordinate and then is in-painted with an official toolbox as the ground truth. As used in \cite{eigen1}\cite{fcrn}, the original frames of size $640\times 480$ is down-sampled by half and center-cropped to $304\times 228$.

\vspace{-1mm}
{\flushleft {\bf KITTI Dataset.}}
Two KITTI datasets are used in this paper: 1) The \textit{Odometry} dataset~\cite{kitti2012} contains 22 sequences with camera and LiDAR measurements. As described in \cite{sparse-to-dense}\cite{cspn}, half of them is used for training and a random subset of 3200 images from the other half is used for evaluation. Following the setting in~\cite{sparse-to-dense}\cite{late-fusion}, the original image is bottom-cropped to $912\times 228$ since there is no depth at the top area;
2) The \textit{Depth Completion} dataset~\cite{kitti2017} focuses on recovering dense depth from a single LiDAR sensor. It contains over 93 thousands depth maps with corresponding raw LiDAR scans and RGB images, where the dense ground truth is collected by accumulating LiDAR measurements from nearby frames in the video sequences with outliers removed~\cite{kitti2017}.

\vspace{-1mm}
{\flushleft {\bf Evaluation Metrics.}}
We adopt standard metrics in depth estimation. Given the ground truth depth and predicted depth, we use the following metrics: Root Mean Square Error (RMSE), Mean Absolute Error (MAE), Mean Absolute Relative error (MRE), and $\delta$ Threshold \cite{eigen1}. Please note that only for $\delta$ Threshold, it is larger the better.
%

\subsection{Overall Performance}
\label{ssec:baselines}

In Table~\ref{tab:baselines-nyu} and Table~\ref{tab:baselines-kitti}, we demonstrate that the performance of both depth estimation and reconstruction baselines are improved by using the proposed PnP module without the need of model re-training on the NYU and KITTI datasets.
In each table, we group the methods according to the input type of a model, denoted as RGB image only (\textbf{rgb}), sparse depth only (\textbf{sd}), and RGB image together with sparse depth (\textbf{rgb+sd}). We follow the settings, e.g., number/percentage of sparse samples, as used in each method.
Among the evaluated methods, \cite{sparse-to-dense}\cite{eigen1}\cite{sfm-learner} are tested on the KITTI \textit{Odometry} dataset while the others are evaluated on the KITTI \textit{Depth Completion} dataset.
%
%

\begin{figure}[t!]
\centering
\includegraphics[width=0.45\textwidth]{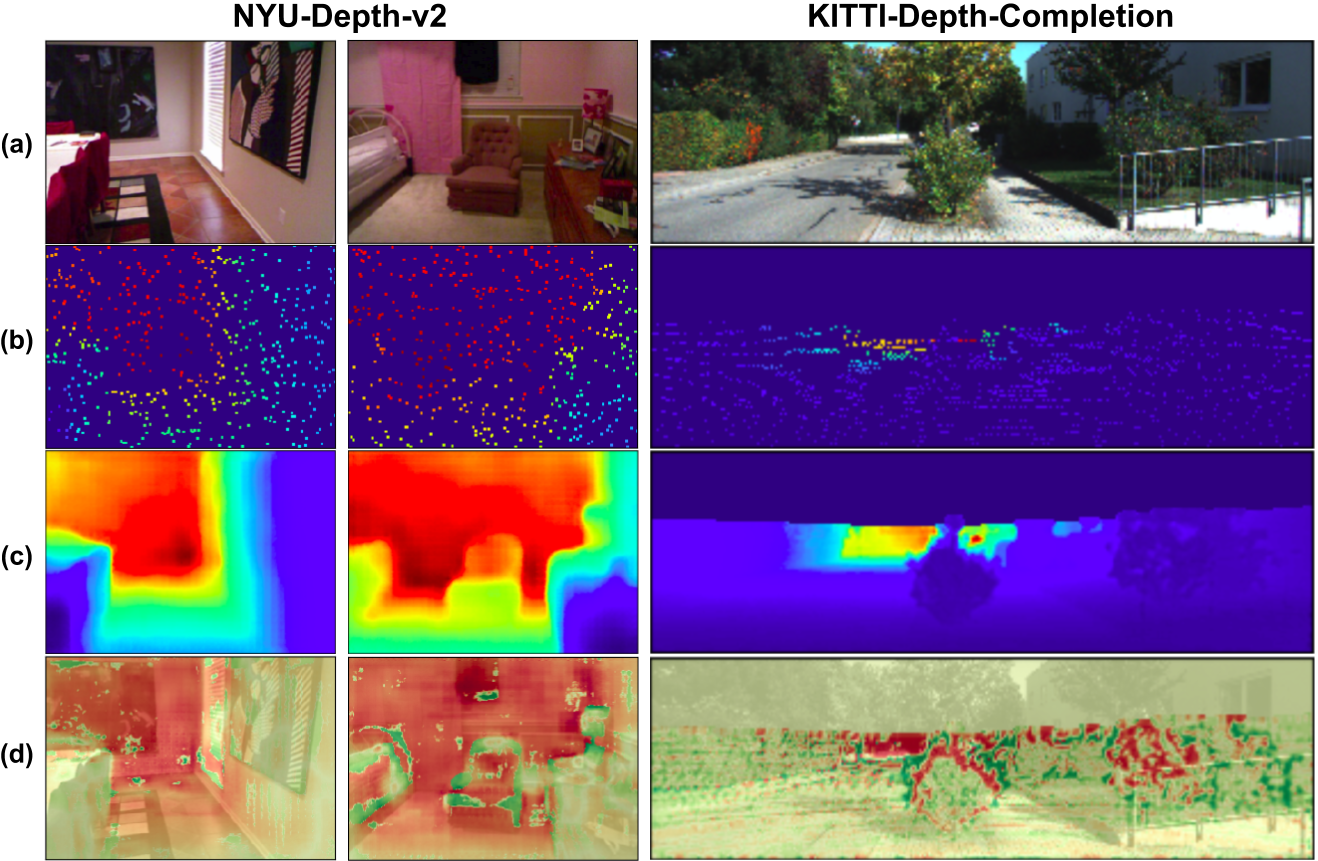}
\vspace{-1mm}
\caption{\small{Qualitative results on NYU and KITTI. Row (a), (b), (c), (d) are RGB image, sparse depth, final prediction with our PnP module, and corresponding improvement map. In (d), the red region indicates improvement and the green one indicates degradation.}}
\label{fig:qualitative_results}
\vspace{-2mm}
\end{figure}

For the \textbf{rgb} case, since both methods~\cite{eigen1}\cite{fcrn} do not use the sparse depth as the guidance in the first place, the performance are thus improved significantly, e.g., larger than 25\% on NYU-Depth-v2 in Table~\ref{tab:baselines-nyu}.
In addition, our method can be generalized to the unsupervised learning-based model~\cite{sfm-learner}, which achieves more than 15\% improvement on KITTI in Table~\ref{tab:baselines-kitti}.
In this setting, we show the benefit of combining the RGB-based depth prediction with the proposed PnP module.
%
%
For the other cases (\textbf{sd} and \textbf{rgb + sd}), although the sparse depth is already utilized in these baseline methods, we show that with the proposed PnP module, the performance can be further improved under various sparsity patterns. We also note that, due to the fact that the NYU and KITTI datasets have different scales of sparse data, we observe relatively larger improvement on NYU than that on KITTI. Some qualitative results are presented in Fig.~\ref{fig:qualitative_results}.
 

\begin{figure}[!t]
\centering
\includegraphics[width=0.5\textwidth]{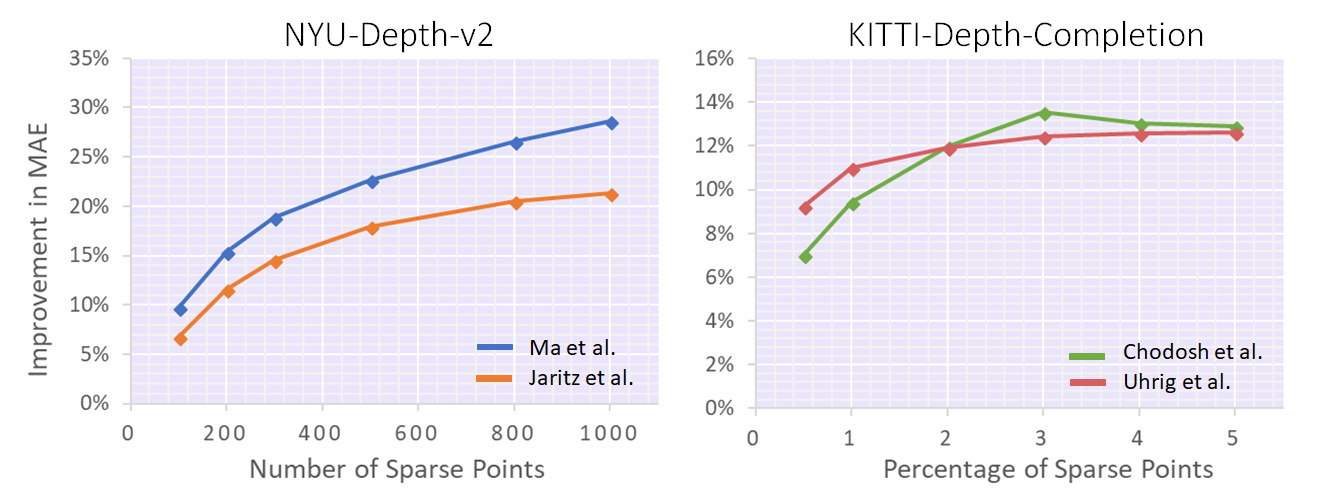}
\vspace{-5mm}
\caption{\small{Improvement with different numbers of sparse points.}}
\label{fig:sparsity_number}
\vspace{-5mm}
\end{figure}

\vspace{-1mm}
\subsection{Sparsity Patterns}

\vspace{-1mm}
{\flushleft {\bf Number/Percentage of Sparse Points.}}
In real-world tasks, different sensors produce various numbers of sparse points.
In Fig.~\ref{fig:sparsity_number}, we show such a study by applying the proposed PnP module to evaluated methods.
On NYU-Depth-v2, we consider \cite{sparse-to-dense} and \cite{late-fusion} as they are designed for using the sparse depth with sparse points, and we follow \cite{sparse-to-dense} to perform uniform sampling. For both baselines, our method provides consistent improvements as the number of sparse points increases.
On KITTI \textit{Depth Completion}, We follow \cite{kitti2017} to sub-sample LiDAR data comparing two baselines~\cite{kitti2017}\cite{deep-compressed-sensing}. Although the improvement degrades a bit for \cite{deep-compressed-sensing}, it is of great interest to observe performance gain using LiDARs, where such measurements often contain missing data.

\vspace{-1mm}
{\flushleft {\bf Sparse Points from LiDARs.}}
In Table~\ref{tab:sparsity_pattern}, we further show that our method can be applied to practical applications, e.g., autonomous vehicles, which usually acquire the raw depth from LiDARs. We simulate four types of Velodyne LiDAR (i.e., \textbf{VLP-16}, \textbf{VLP-32C}, \textbf{HDL-32E}, and \textbf{HDL-64E}) which are distinct on their field of view (\textbf{FoV}) and vertical angular resolution (\textbf{vRes}), with additional random noise on rotation.
We evaluate \cite{sparse-to-dense} on the SYNTHIA dataset~\cite{synthia} since it is the only outdoor dataset with dense ground truth that is required by LiDAR sampling. 
In Table~\ref{tab:sparsity_pattern}, we present the \textbf{MAE} without using our PnP module and its relative improvement \textbf{MAE+}. Among these four types, \textbf{VLP-32C} obtains the largest improvement since it has the largest FoV and number of sparse samples, followed by \textbf{HDL-32E} which has a similar FoV and lower vRes (lower number means higher resolution). Interesting, \textbf{HDL-32E} with fewer sparse points and less density (vRes) performs better than \textbf{HDL-64E}, suggesting that the spatial distribution of sparse points (FoV) is more important than the density of sparse points (vRes) for depth reconstruction.
%

\begin{table}[!t]
\centering
\caption{\small{Improvement with different types of LiDAR simulations on SYNTHIA. FoV and vRes are field of view and vertical resolution (lower number means higher resolution). MAE and MAE+ are the performance without PnP module and the relative improvement in MAE using our method, respectively.}}
\vspace{-1mm}
\begin{tabular}{l r r r r r}
\hline
Source & \%Samples & FoV & vRes & MAE~$\shortdownarrow$ & MAE+~$\shortuparrow$ \\ \hline
VLP-16  & $5.3$ & $30$ & $2.0$ & $1.274$ & $9.3\%$ \\
VLP-32C & $28.8$ & $40$ & $0.3$ & $1.075$ & $11.8\%$ \\
HDL-32E & $10.3$ & $41$ & $1.3$ & $1.109$ & $11.1\%$ \\
HDL-64E & $18.7$ & $27$ & $0.4$ & $1.236$ & $8.9\%$ \\
\hline
\end{tabular}
\small
\label{tab:sparsity_pattern}
\vspace{-2mm}
\end{table}

\begin{figure}[t!]
\centering
\includegraphics[width=0.43\textwidth]{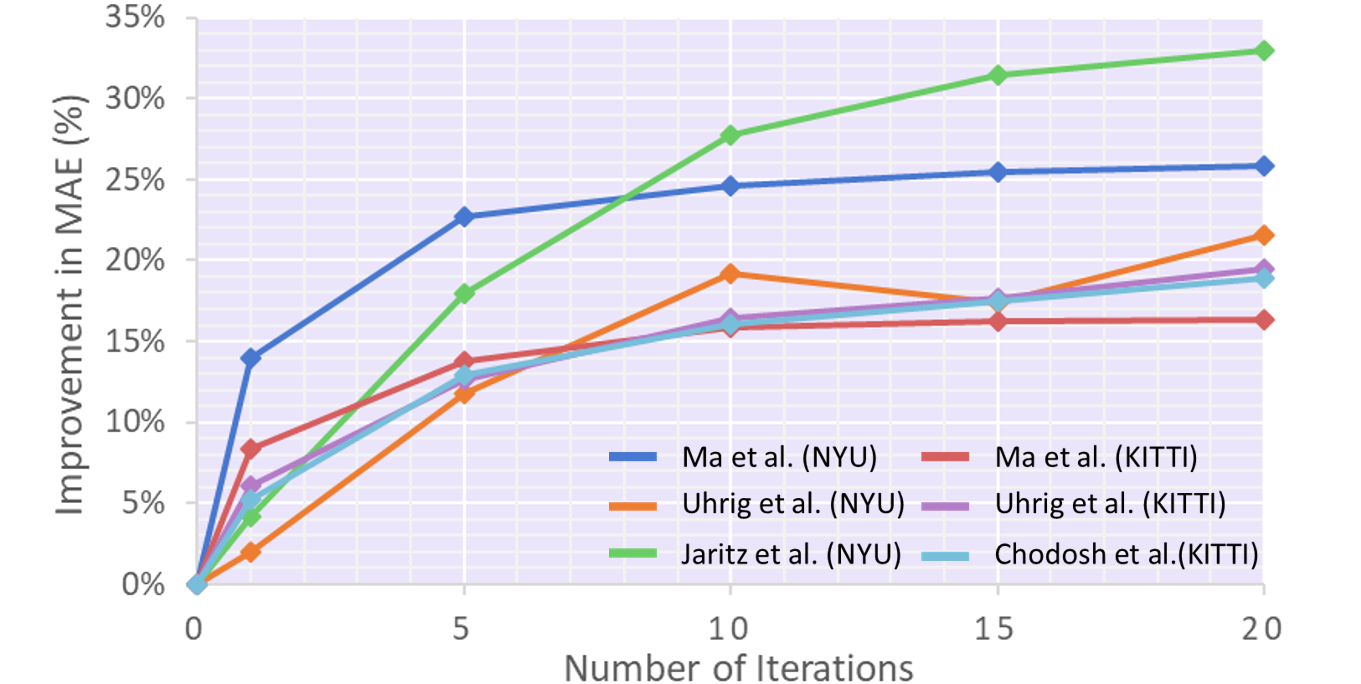}
\vspace{-1mm}
\caption{\small{Improvement under multiple iterations.}}
\label{fig:iter}
\vspace{-2mm}
\end{figure}

\begin{figure}[t!]
\centering
\includegraphics[width=0.44\textwidth]{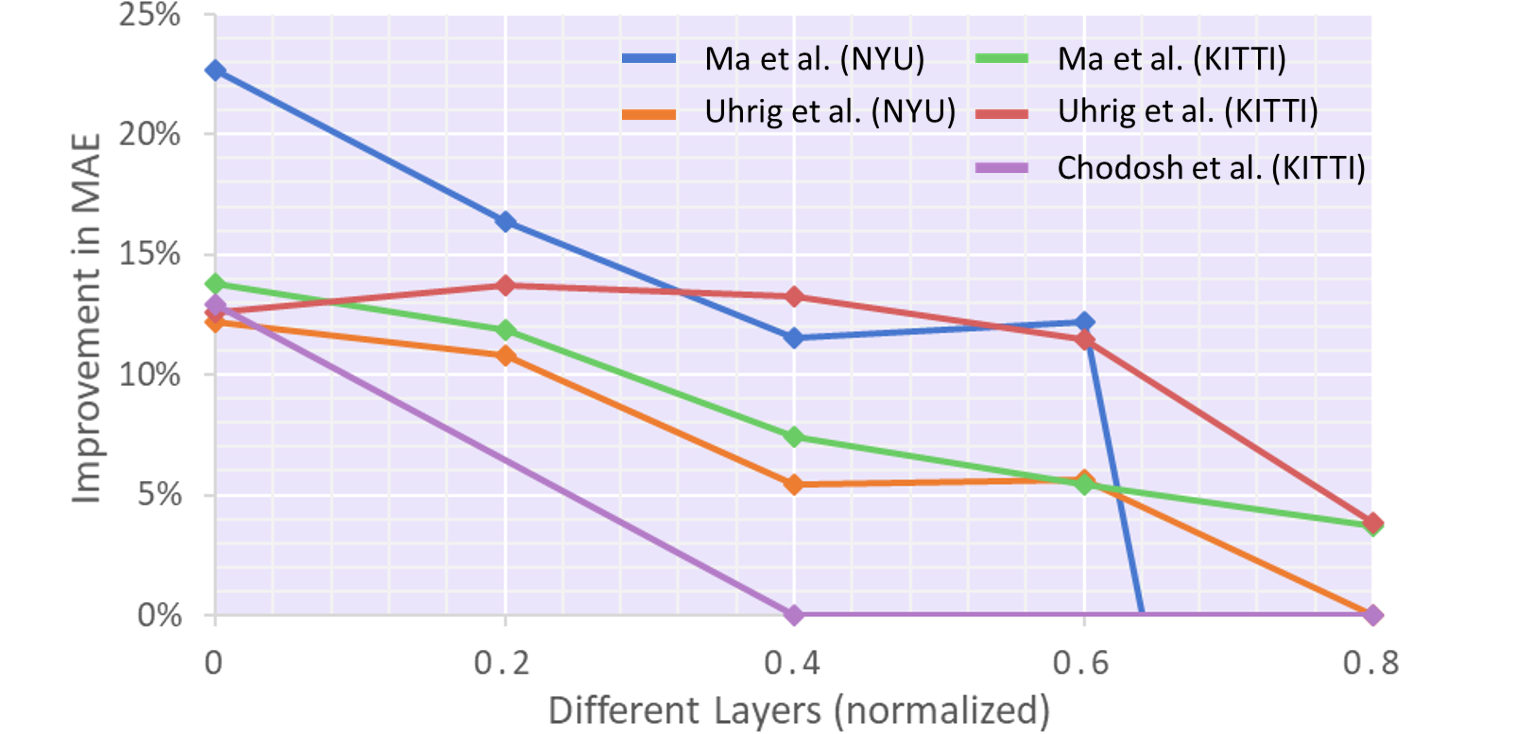}
\vspace{-1mm}
\caption{\small{Improvement under different layers during back-propagation. The layer number is normalized for better illustration.}}
\label{fig:layer}
\vspace{-5mm}
\end{figure}

\vspace{-1mm}
\subsection{Ablation Study}
\vspace{-1mm}
{\flushleft {\bf Effects of Iterative Optimization.}}
In Fig.~\ref{fig:iter}, we demonstrate that our method is capable of gradually propagating the information of sparse depth over multiple iterations, where the improvement are increased as the number of iterations grows. Nevertheless, since more iterations stand for longer inference time and the improvement gradually saturates when the number of iterations is sufficiently large, we would not prefer to have extremely many iterations in practical usage considering the trade-off between runtime and accuracy.


\vspace{-1mm}
{\flushleft {\bf Back-propagation to Different Layers.}}
Here, we investigate different choices of intermediate feature map $z$. In this experiment, we fix the number of iterations and learning rate $\alpha$ as 5 and 0.01, respectively. In Fig.~\ref{fig:layer}, the horizontal axis starts from shallow to deep layers and is normalized since the number of layers in each model is different.
For instance, we set the first layer of the decoder in Ma \textit{et al.} as $0$. For Uhrig \textit{et al.} and Chodosh \textit{et al.}, which are not the encoder-decoder architecture and only contain 6 and 3 layers, respectively, we start from the first layer of the entire network as the intermediate feature $z$ and proceed.

In Fig.~\ref{fig:layer}, choosing $z$ further from the output layer (i.e., smaller number in the horizontal axis) yields better results since it produces a larger influential field (as described in Sec.~\ref{ssec:how_to_use}) and this observation holds for both NYU and KITTI. Furthermore, predictions could be more sensitive when updating the feature map that is closer to the output (e.g., Ma \textit{et al.} on NYU).
In addition, it is worth mentioning that the improvement in Chodosh et al.~\cite{deep-compressed-sensing} saturates quickly. The reason is that this method already shares a similarity to ours in a sense that both aim to optimize a latent representation that is consistent to the given sparse map.

\begin{table}[!t]
\centering
\caption{\small{Inference time of applying PnP to different models.}}
\vspace{-1mm}
\begin{tabular}{l r r r r r}
\hline
Dataset & \multicolumn{2}{c}{NYU} & \multicolumn{2}{c}{KITTI} \\ \hline
Model & Ma \textit{et al.} & Jaritz \textit{et al.} & Uhrig \textit{et al.} & Chodosh \textit{et al.} \\ \hline \hline
Original & 0.0046 & 0.0484 & 0.0311 & 0.0320 \\
w/ PnP & 0.0233 & 0.1593 & 0.1614 & 0.0447 \\
\hline
\end{tabular}
\small
\label{tab:infer_time}
\vspace{-5mm}
\end{table}

\vspace{-1mm}
\subsection{Computational Time}{
\flushleft In Table~\ref{tab:infer_time}, we show the computational time applying our PnP module to different models on both NYU and KITTI. We fix the number of iterations as 5. The inference time mainly depends on the model architecture and the selection of intermediate representation $z$. For instance, performing our PnP module on Chodosh \textit{et al.} brings only slightly computational time overhead, because the model only contains three layers and $z$ is chosen much further from the output layer.
}

\section{Conclusions}

In this paper, we propose a plug-and-play (PnP) module for improving depth prediction given arbitrary patterns of sparse depths, that can be applied to any differentiable model without re-training.
We validate our PnP module using several state-of-the-art depth prediction methods with various input types on indoor (NYU) and outdoor (KITTI) datasets, achieving consistent improvements. In addition, we synthesize various types of LiDARs and verify the general applicability of our method.
%
Finally, comprehensive analysis is carried out to provide a simple rule of thumb for applying our PnP module.

\vspace{-2mm}
{\flushleft {\bf Acknowledgement.}} This project is supported by MOST Joint Research Center for AI Technology and All Vista Healthcare with program MOST 108-2634-F-007-006 and MOST-108-2636-E-009-001.

\bibliographystyle{IEEEtran}
\bibliography{ref}

\begin{thebibliography}{10}
\providecommand{\url}[1]{#1}
\csname url@rmstyle\endcsname
\providecommand{\newblock}{\relax}
\providecommand{\bibinfo}[2]{#2}
\providecommand\BIBentrySTDinterwordspacing{\spaceskip=0pt\relax}
\providecommand\BIBentryALTinterwordstretchfactor{4}
\providecommand\BIBentryALTinterwordspacing{\spaceskip=\fontdimen2\font plus
\BIBentryALTinterwordstretchfactor\fontdimen3\font minus
  \fontdimen4\font\relax}
\providecommand\BIBforeignlanguage[2]{{%
\expandafter\ifx\csname l@#1\endcsname\relax
\typeout{** WARNING: IEEEtran.bst: No hyphenation pattern has been}%
\typeout{** loaded for the language `#1'. Using the pattern for}%
\typeout{** the default language instead.}%
\else
\language=\csname l@#1\endcsname
\fi
#2}}

\bibitem{kitti2017}
J.~Uhrig, N.~Schneider, L.~Schneider, U.~Franke, T.~Brox, and A.~Geiger,
  ``Sparsity invariant cnns,'' in \emph{International Conference on 3D Vision
  (3DV)}, 2017.

\bibitem{sparse-to-dense}
F.~Ma and S.~Karaman, ``Sparse-to-dense: Depth prediction from sparse depth
  samples and a single image,'' in \emph{International Conference on Robotics
  and Automation (ICRA)}, 2018.

\bibitem{late-fusion}
M.~Jaritz, R.~De~Charette, E.~Wirbel, X.~Perrotton, and F.~Nashashibi, ``Sparse
  and dense data with cnns: Depth completion and semantic segmentation,''
  \emph{arXiv preprint arXiv:1808.00769}, 2018.

\bibitem{magic}
Z.~Chen, V.~Badrinarayanan, G.~Drozdov, and A.~Rabinovich, ``Estimating depth
  from rgb and sparse sensing,'' \emph{arXiv preprint arXiv:1804.02771}, 2018.

\bibitem{deep-compressed-sensing}
N.~Chodosh, C.~Wang, and S.~Lucey, ``Deep convolutional compressed sensing for
  lidar depth completion,'' \emph{arXiv preprint arXiv:1803.08949}, 2018.

\bibitem{cspn}
R.~Y. Xinjing~Cheng, Peng~Wang, ``Depth estimation via affinity learned with
  convolutional spatial propagation network,'' in \emph{European Conference on
  Computer Vision (ECCV)}, 2018.

\bibitem{deep-visual}
J.~Yosinski, J.~Clune, A.~Nguyen, T.~Fuchs, and H.~Lipson, ``Understanding
  neural networks through deep visualization,'' in \emph{Deep Learning
  Workshop, International Conference on Machine Learning (ICML)}, 2015.

\bibitem{fast-grad}
A.~Kurakin, I.~Goodfellow, and S.~Bengio, ``Adversarial examples in the
  physical world,'' \emph{arXiv preprint arXiv:1607.02533}, 2016.

\bibitem{nyu-v2}
P.~K. Nathan~Silberman, Derek~Hoiem and R.~Fergus, ``Indoor segmentation and
  support inference from rgbd images,'' in \emph{European Conference on
  Computer Vision (ECCV)}, 2012.

\bibitem{kitti2012}
A.~Geiger, P.~Lenz, and R.~Urtasun, ``Are we ready for autonomous driving? the
  kitti vision benchmark suite,'' in \emph{Conference on Computer Vision and
  Pattern Recognition (CVPR)}, 2012.

\bibitem{eigen1}
D.~Eigen, C.~Puhrsch, and R.~Fergus, ``Depth map prediction from a single image
  using a multi-scale deep network,'' in \emph{Advances in neural information
  processing systems}, 2014, pp. 2366--2374.

\bibitem{eigen2}
D.~Eigen and R.~Fergus, ``Predicting depth, surface normals and semantic labels
  with a common multi-scale convolutional architecture,'' in \emph{Proceedings
  of the IEEE International Conference on Computer Vision}, 2015, pp.
  2650--2658.

\bibitem{conv-neural-field}
F.~Liu, C.~Shen, and G.~Lin, ``Deep convolutional neural fields for depth
  estimation from a single image,'' in \emph{Proceedings of the IEEE Conference
  on Computer Vision and Pattern Recognition}, 2015, pp. 5162--5170.

\bibitem{fcrn}
I.~Laina, C.~Rupprecht, V.~Belagiannis, F.~Tombari, and N.~Navab, ``Deeper
  depth prediction with fully convolutional residual networks,'' in \emph{3D
  Vision (3DV), 2016 Fourth International Conference on}.\hskip 1em plus 0.5em
  minus 0.4em\relax IEEE, 2016, pp. 239--248.

\bibitem{dorn}
H.~Fu, M.~Gong, C.~Wang, K.~Batmanghelich, and D.~Tao, ``Deep ordinal
  regression network for monocular depth estimation,'' in \emph{Proceedings of
  the IEEE Conference on Computer Vision and Pattern Recognition}, 2018, pp.
  2002--2011.

\bibitem{hawe-wavelet}
S.~Hawe, M.~Kleinsteuber, and K.~Diepold, ``Dense disparity maps from sparse
  disparity measurements,'' in \emph{13th International Conference on Computer
  Vision}, 2011.

\bibitem{liu-wavelet}
L.-K. Liu, S.~H. Chan, and T.~Q. Nguyen, ``Depth reconstruction from sparse
  samples: Representation, algorithm, and sampling,'' \emph{IEEE Transactions
  on Image Processing}, vol.~24, no.~6, pp. 1983--1996, 2015.

\bibitem{tgv-anisotropic}
D.~Ferstl, C.~Reinbacher, R.~Ranftl, M.~R{\"u}ther, and H.~Bischof, ``Image
  guided depth upsampling using anisotropic total generalized variation,'' in
  \emph{Proceedings of the IEEE International Conference on Computer Vision},
  2013, pp. 993--1000.

\bibitem{parse-geometry}
Y.~Liao, L.~Huang, Y.~Wang, S.~Kodagoda, Y.~Yu, and Y.~Liu, ``Parse geometry
  from a line: Monocular depth estimation with partial laser observation,'' in
  \emph{Robotics and Automation (ICRA), 2017 IEEE International Conference
  on}.\hskip 1em plus 0.5em minus 0.4em\relax IEEE, 2017, pp. 5059--5066.

\bibitem{deep-depth-completion}
Y.~Zhang and T.~Funkhouser, ``Deep depth completion of a single rgb-d image,''
  in \emph{Proceedings of the IEEE Conference on Computer Vision and Pattern
  Recognition}, 2018.

\bibitem{cao}
Y.~Cao, Z.~Wu, and C.~Shen, ``Estimating depth from monocular images as
  classification using deep fully convolutional residual networks,'' \emph{IEEE
  Transactions on Circuits and Systems for Video Technology}, 2017.

\bibitem{back-prop}
D.~E. Rumelhart, G.~E. Hinton, and R.~J. Williams, ``Learning internal
  representations by error propagation,'' in \emph{Parallel Distributed
  Processing: Explorations in the Microstructure of Cognition, {V}olume 1:
  {F}oundations}, 1986.

\bibitem{adam}
D.~P. Kingma and J.~Ba, ``Adam: A method for stochastic optimization,'' in
  \emph{International Conference for Learning Representations}, 2014.

\bibitem{sfm-learner}
T.~Zhou, M.~Brown, N.~Snavely, and D.~G. Lowe, ``Unsupervised learning of depth
  and ego-motion from video,'' in \emph{CVPR}, 2017.

\bibitem{synthia}
G.~Ros, L.~Sellart, J.~Materzynska, D.~Vazquez, and A.~M. Lopez, ``The synthia
  dataset: A large collection of synthetic images for semantic segmentation of
  urban scenes,'' in \emph{The IEEE Conference on Computer Vision and Pattern
  Recognition (CVPR)}, June 2016.

\end{thebibliography}
\normalsize


\end{document}